\documentclass[a4paper,fleqn,usenatbib,onecolumn]{mnras}
\usepackage[T1]{fontenc}
\usepackage{ae,aecompl}
\usepackage{graphicx}	
\usepackage{amsmath}	
\usepackage{amssymb}	
\title[Twisted magnetosphere in the exterior of a neutron star]
{Twisted magnetosphere with quadrupolar fields
in the exterior of a neutron star}
\author[Y. Kojima]{Yasufumi Kojima
\thanks{%
E-mail: ykojima-phys@hiroshima-u.ac.jp}\\ 
Department of Physics, Hiroshima University, Higashi-Hiroshima, Hiroshima 
739-8526, Japan}
\begin{document}
 \label{firstpage}
 \pagerange{\pageref{firstpage}--\pageref{lastpage}}
 \maketitle
%
\begin{abstract}
The magnetar magnetosphere is gradually twisted by shearing from footpoint 
motion, and stored magnetic energy also increases at the same time.
When a state exceeds a threshold, flares/outbursts manifest themselves as 
a result of a catastrophic transition.
Axisymmetric static solutions for a relativistic 
force-free magnetosphere with dipole--quadrupole
mixed fields at the surface have been calculated.
The quadrupole component represents a kind of 
magnetic-field irregularity at a small scale.
Locally twisted models are constructed by limiting current flow regions, 
where the small part originates from a dipole--quadrupole mixture.
The energy along a sequence of equilibria increases and becomes sufficient
to open the magnetic field in some models.
In energetically metastable states,
a magnetic flux rope is formed in the vicinity of the star. 
The excess energy may be ejected as a magnetar flare/outburst.
The general relativistic gravity is sufficient to confine
the flux rope and to store huge magnetic energy, and
the mechanism is also discussed.
\end{abstract}
\begin{keywords}
stars: magnetars -- stars: neutron -- stars: magnetic fields
\end{keywords}

\section{Introduction}

%
Magnetars are a class of neutron stars with extremely strong magnetic fields.
Their activities, such as sudden flares/bursts and persistent X-ray emissions, 
are powered by magnetic fields
\citep[e.g.,][for a review]{2015RPPh...78k6901T,2017ARA&A..55..261K}.
The interior magnetic field structure is never permanently fixed, but decays 
into heat on a secular timescale.
The ohmic dissipation timescale for a global field structure is too long
among possible mechanisms for this evolution.
Therefore, other mechanisms
e.g., Hall drift, ambipolar diffusion, or their combinations, are relevant.
For example, the magnetic field drifts to be localized to a certain region in which ohmic dissipation is effectively enhanced.
The evolutionary timescales for Hall drift and ambipolar diffusion 
are proportional to
$B$ and $B^2$, where $B$ is typical magnetic-field strength
\citep[][]{1992ApJ...395..250G}.
These mechanisms may therefore be relevant to neutron stars with strong fields, 
e.g., $ B > 10^{14}$G. 
The magnetic fields for typical pulsars with $ B \sim 10^{12}$G
are almost unchanged by them.
Among the other factors, magnetic field evolution has been extensively
simulated under Hall drift in neutron-star crusts
\citep[e.g.,][]{2004MNRAS.347.1273H,2012MNRAS.421.2722K,
2013MNRAS.434..123V,2014MNRAS.438.1618G,2015PhRvL.114s1101W}. 
For example, \citet{2014MNRAS.444.3198G} calculated the evolution 
in axially symmetric configuration, together with thermal evolution. 
Their magneto-thermal simulation shows a remarkable feature --
a magnetic spot caused by an initially confined toroidal component
that manifests itself on the surface around $10^4$ years.
Furthermore, \citet{2016PNAS..113.3944G}
extended the Hall-magnetohydrodynamic simulation to a three-dimensional 
configuration and showed intense magnetic features at small-scales are driven
by non-axially symmetric instabilities.
Thus, higher multi-poles are likely to be produced via nonlinear coupling 
between poloidal and toroidal fields in the crust.
There remain subtle problems concerning dependence of
the initial magnetic geometries and the external fields. 
Such evolutions have been calculated in most previous research under the assumption that the exterior is in vacuum. 
Hall drift of magnetic field lines build up stresses in the crust, and 
shearing motion at the stellar surface twists the exterior magnetic field, 
irrespective of whether the neutron star crust responds to the magnetic stress 
elastically or plastically. 
The twisted structure is supported by flowing currents.
When the state exceeds a threshold, the energy is abruptly released 
on a much shorter dynamical timescale, leading to energetic flares.
It is important to take into account
the effect of the twisted magnetosphere instead of vacuum on the evolution.
Recently, \citet{2017MNRAS.472.3914A}
modeled this evolution by taking into account the force-free magnetosphere.
Their evolution model shows that there is no equilibrium solution for the 
exterior magnetosphere on timescales of the order of thousands of years.  
The breakdown of models at that time suggests an outburst.
Subsequent evolution after the burst is not followed, as 
the rearrangement of magnetic fields cannot be calculated.
Their initial configuration is simplified as a bipolar one, and
there is not enough time for a transition to higher multi-pole states.
Static equilibrium solutions are helpful for understanding the 
magnetar magnetosphere.
The force-free approximation may be applicable, since
magnetospheres are filled by low-density plasma.
Important elements involved in magnetar magnetospheres have been studied
in the static approach.
For example, twisted force-free magnetospheres around magnetars
have been numerically constructed as a part of entire magnetic field structures 
from stellar cores to exteriors
\citep{2014MNRAS.437....2G,2014MNRAS.445.2777F,2015MNRAS.447.2821P,
2017MNRAS.470.2469P}.
 \citet{2016MNRAS.462.1894A}
studied the effect of a surrounding current-free region on
a twisted magnetosphere. 
One of the interesting properties found in some numerical results
is the formation of a magnetic flux rope, 
an axially symmetric torus in the vicinity of the stellar surface, 
when the magnetic field is highly twisted.
The topology in the presence of the rope
differs from that of the current-free potential field. 
There are multiple force-free solutions under 
fixed boundary conditions, and
a higher state containing the flux rope is obtained by changing the
iteration scheme to solve a non-linear Grad--Shafranov equation
 \citep{2018MNRAS.474..625A}.
The spacetime outside the magnetar is assumed to be flat
except in a few studies \citep{2015MNRAS.447.2821P,2017MNRAS.470.2469P,
2017MNRAS.468.2011K,2018MNRAS.475.5290K}.
Treatment in flat spacetime seems to be reasonable as the
lowest order approximation; a priori, the correction is expected 
to be not so large, since the relativistic factor is of order 
$G_{\rm N}M/(Rc^2) \sim 0.2-0.3$ in neutron stars.
In our previous studies \citep{2017MNRAS.468.2011K,2018MNRAS.475.5290K}, 
however, we found that general relativistic effects are noteworthy.
The maximum energy stored in a current-flowing relativistic magnetosphere
increases by 132 percent from the current free dipole one.
This contrasts with the maximum excess energy of only 34 percent 
in a non-relativistic model.
This large increase in a relativistic model is related to the formation 
of a flux rope, 
an axially symmetric torus in the vicinity of the stellar surface, 
when the magnetic field structure is highly twisted.
Curved spacetime helps to confine the torus.
It is reasonable to infer that near the magnetar surface, the magnetic 
field geometry involves higher multi-poles.
Active regions manifest themselves as multi-polar regions.
Therefore, our previous calculations are further extended to
examine the effect of a higher multi-pole in this study.
By considering a mixed dipole and quadrupole configuration
at the surface, a locally twisted model is constructed.
Relevant amounts of energy should be reduced by
limiting the current flow region in the vicinity of a neutron star. 
This energy that comes from non-potential magnetic fields 
is available for rapid release through a variety
of mechanisms that may involve instabilities, 
loss of equilibrium, and/or reconnection. 
When a detached magnetic flux is ejected, the structure
becomes temporarily open.
The structural change to an open field configuration is therefore 
impossible with respect to energy, 
from a state with energy stored in the magnetosphere
exceeding open field energy. 
Our concern is whether or not the stored energy exceeds open field energy
in the presence of multi-pole fields.
  This study is organized as follows. We briefly discuss our model and 
relevant equations for a non-rotating force-free magnetosphere
in a Schwarzschild spacetime in Section 2.
We then numerically solve the so-called Grad--Shafranov equation
assuming that the current function is given by a 
simple power-law model.
In Section 3, the results are given 
for comparison with those obtained in flat spacetime.
Finally, Section 4 summarizes and discusses the implications of our results.
We use geometrical units of $c=G_{\rm N}=1$.

\section{Equations}
  \subsection{Force-free magnetosphere in a curved spacetime}
In this section, we briefly summarize our formalism
by vector analysis in curvilinear coordinates
\citep[see][for details]{2017MNRAS.468.2011K}.
We consider the static magnetic configuration
for the exterior of a non-rotating compact object with a mass $M$
and radius $R$.
The spacetime is described by the Schwarzschild metric for $R>2M$.
The magnetic field for the axially symmetric case is described by 
two functions $G$ and $S$: 
\begin{equation}
\vec{B}=\vec{\nabla}\times \left(
\frac{G}{\varpi}\vec{e}_{\hat{\phi}}
\right)
+\frac{S}{\alpha \varpi}\vec{e}_{\hat{\phi}}
=
\frac{\vec\nabla{G}\times\vec{e}_{\hat{\phi}}}{\varpi}
+\frac{S}{\alpha \varpi}\vec{e}_{\hat{\phi}},
\label{eqnDefBB}
\end{equation}
and its components can be written as
\begin{equation}
[B_{\hat{r}},B_{\hat{\theta}}, B_{\hat{\phi}}]
=\left[\frac{G,_\theta}{r \varpi},
~
 -\frac{\alpha G,_r}{\varpi},
~
 \frac{S}{\alpha \varpi}\right],
\label{Bcomp}
\end{equation}
where $\alpha=(1-2M/r)^{1/2}$ and $ \varpi = r \sin \theta$.
The magnetic flux function $G$ describes
poloidal magnetic fields, and function $S$
describes poloidal current flow as 
 $4 \pi \alpha \vec{j}_{p}=\vec{\nabla}\times (\alpha \vec{B})$
 $=\vec\nabla{S}\times\vec{e}_{\hat{\phi}}/\varpi $.
In a force-free approximation, the current flows along magnetic field lines,
and the current function $S$ should be a function of $G$.
We consider a specific power-law model with a positive constant $\gamma$:
\begin{equation}
S=\left(\frac{\gamma}{3} \right)^{1/2}G^{3} ,
\label{powerlaw}
\end{equation}
such that the current is simply given by 
\begin{equation}
4\pi \alpha {\vec j} = (3\gamma)^{1/2} G^2 {\vec B} .
\label{eqn:current}
\end{equation}
In this model, the current always flows in 
the same direction as the magnetic field.
The current distribution is weighted in favor of large $|G|$ values, such that
the current is likely to be localized. 
The global structure of the magnetic field is determined by
solving the azimuthal component of the Biot--Savart equation
with current $ j_{\hat \phi} $:
\begin{equation}
\frac{\varpi}{\alpha}\vec{\nabla} \cdot
\left(\frac{\alpha}{\varpi^2} \vec{\nabla} G \right)
= - 4\pi j_{\hat \phi} .
\label{eqn:FF}
\end{equation}
The left hand side of this equation is explicitly written as
\begin{equation}
\frac{1}{\varpi}
\left[ \frac{\partial}{\partial r}
  \left( \alpha^2\frac{\partial G}{\partial r} \right)
  +\frac{ \sin\theta}{r^2} \frac{\partial}{\partial \theta}
  \left(\frac{1}{\sin \theta}\frac{\partial G}{\partial \theta}\right) 
\right] ,
\end{equation}
and the source term of eq.(\ref{eqn:FF})
is simply reduced to $ -\gamma G^5/(\alpha^{2} \varpi) $
[power-law model with $n=5$ in 
\citep{2004ApJ...606.1210F,2017MNRAS.468.2011K}].
It is instructive to rewrite eq.(\ref{eqn:FF}) as
\begin{equation}
\varpi \vec{\nabla} \cdot
\left(\frac{1}{\varpi^2} \vec{\nabla} G \right)
= - 4\pi (j_{\hat \phi} + j_{g}) ,
\label{eqn:FFNR}
\end{equation}
where 
\begin{equation}
4 \pi j_{g}  \equiv  \frac{1}{\varpi} \vec{\nabla} \ln \alpha
      \cdot \vec{\nabla} G 
            =  -\frac{1}{  \varpi }\vec{g} \cdot \vec{\nabla} G .
\label{eqn:GRj}
\end{equation}
A new term $ j_{g}$ in eq.(\ref{eqn:FFNR}) arises from gravitational 
acceleration $\vec{g}= -\vec{\nabla} \ln \alpha$ in general relativity.
It is ignored in treatment for flat spacetime.
Indeed, the magnitude of $j_{g}$ is not so large, as
$ M/R \approx 0.2$ times a true current. 
However, the 'general-relativity-induced current' $j_{g}$ 
plays an important role on the confinement of a flux rope,
as will be discussed in the next section.

  \subsection{Mixture of dipole and quadrupole}
\begin{figure}
\centering
\includegraphics[scale=1.0]{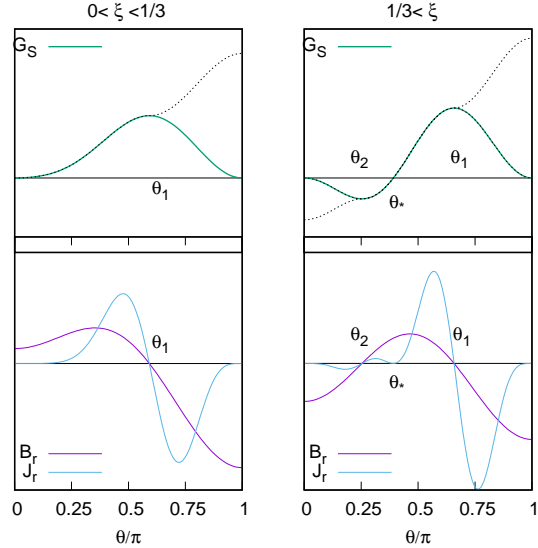}
\caption{
Magnetic function $G_{S}$ (top panel), radial component of
magnetic field $B_{\hat r}$,
and current  $j_{\hat r}$ (bottom panel) at the surface
as a function of polar angle $\theta$.
The vertical axis is arbitrary scaled.
The left panels are for a dipole-dominated field
with a parameter $0 < \xi < 1/3$, 
and the right panels are 
for a quadrupole-dominated field with $1/3 < \xi $.
The function $G_{S}$ becomes zero at a certain angle $\theta _*$
$(0< \theta _* <\pi)$ in the latter.
Zeros of $B_{\hat r}=0$ at the surface
are denoted in the figure as $\theta = \theta_{n}~(n=1,2)$.
In the top panels, a dotted line corresponds to
a monopolar configuration.
}
\label{fig:0}
\end{figure}

%
It is useful to show analytic solutions in vacuum,
for eq.(\ref{eqn:FF}) with ${\vec j}=0$.
The magnetic function $G$ is expanded using Legendre polynomials 
$P_{l}(\theta)$:
\begin{equation}
G(r,\theta )=-\sum_{l \ge 1}g_{l}(r)\sin \theta 
\frac{d P_{l}(\theta )}{d\theta }.
   \label{Sexpd.eqn}
\end{equation}
The radial functions $g_{l}$ are for example given by
\cite[e.g.][for higher $l$]{2017MNRAS.472.3304P} 
\begin{eqnarray}
 \label{eqnpuredip}
g_{1} &=& - \frac{3 B_{0} R^3 r^2}{8 M^3}\left[
\ln\left(1-\frac{2M}{r}\right) 
+\frac{2M}{r}+\frac{2M^2}{r^2} \right]
\\
&\approx &
\frac{ B_{0} R^3 }{r}\left[
1+\frac{3M}{2r}+\frac{12M^2}{5r^2} +\cdots \right],
~~~~~ (M/r \ll 1)
\nonumber
\end{eqnarray}
\begin{eqnarray}
 \label{eqnpurequad}
g_{2} &=& - \frac{ 5 B_{0} R^4 r^3}{2 M^5}\left[
\left(1-\frac{3M}{2r}\right) 
\ln\left(1-\frac{2M}{r}\right) 
+\frac{M}{r}
\left(2-\frac{M}{r}-\frac{M^2}{3r^2}
\right) 
\right]
\\
&\approx &
\frac{ B_{0}R^4}{r^2}\left[
1+\frac{8M}{3r}+\frac{40M^2}{7r^2} +\cdots
\right],
~~~~~ (M/r \ll 1)
\nonumber
\end{eqnarray}
where $ B_0 $ is typical field strength and 
$R$ is the stellar surface radius.
In these expressions, the first expression is an exact solution,
and the second is its approximation under the weak gravity regime
$M/r \ll 1$.
Multi-pole moments are generally defined by the asymptotic form approaching infinity.
For example, the magnetic dipole moment $\mu$ is given by $ \mu = B_0 R^{3}$.
The surface magnetic field strength $B_{p}$ at magnetic polar cap 
is simply given by
$B_{p}=2B_{0}=2\mu/ R^{3}$ in a flat spacetime model.
However, it should be noted that 
additional relativistic corrections are needed to connect 
the surface field with $B_{0}$ in a curved spacetime model,
and that the difference between $g_{l}$ and a simple 
power-law solution ($\propto r^{-l}$)
in a flat spacetime increases with $l$.
Accordingly, the relativistic correction is more important in higher multi-poles.
The function $g_{2}$ in eq.(\ref{eqnpurequad})
is also normalized by the same constant $B_0$. 
We solve the non-linear equation (\ref{eqn:FF}) 
using a numerical method described in \cite{2017MNRAS.468.2011K}.
Here, we discuss the boundary conditions.
At the polar axis, the magnetic function $ G$ should satisfy the 
regularity condition, that is, $ G=0$ at $\theta =0$ and $ \pi$.
At asymptotic infinity ($r \to \infty $),
the function should decrease as $G \propto r^{-1} $. 
At the stellar surface $r=R$, the function 
$G_{S}(\theta ) \equiv G(R,\theta ) $ is assumed to be a mixture of 
dipole ($l=1$) and quadrupole ($l=2$) fields with a ratio $a_2$
\footnote{We consider $a_2 >0$ only, since
 $a_2 <0$ corresponds to north-south inversion 
with respect to $\theta$. 
}.
The angular dependence is set by
\begin{eqnarray}
G_{S}(\theta ) & =&- \left( g_{1}(R) \sin \theta 
\frac{d P_{1}(\theta )}{d\theta } -
 a_{2} g_{2}(R) \sin \theta 
\frac{d P_{2}(\theta )}{d\theta } \right) 
\nonumber \\
 & =&  [ g_{1}(R) - 3a_{2}g_{2}(R) \cos\theta ]\sin^2\theta ,
  \label{SurfG.eqn}
\end{eqnarray}
where $g_{l}(R)~ (l=1,2)$ given in eqs. (\ref{eqnpuredip})-(\ref{eqnpurequad})
are incorporated to determine the boundary value.
For a fixed ratio $a_2$ and relativistic factor $M/R$,
the solutions in vacuum are expressed simply by a sum of
eqs. (\ref{eqnpuredip})-(\ref{eqnpurequad}).
The potential field configuration is used as a reference, and 
twisted magnetospheres are constructed 
by including toroidal magnetic field $B_{\hat \phi}$. 
Figure \ref{fig:0} shows $G_{S}(\theta ) $ in eq. (\ref{SurfG.eqn}).
There are two cases that depend on a parameter $\xi  \equiv a_2g_2/g_1$,
which is simply reduced to the ratio $a_{2}$ in a flat spacetime.
One is a dipole-like configuration when $0 \le  \xi < 1/3 $, 
and the other is a quadrupole-like one when  $  \xi \ge 1/3 $.
In the latter, $G_{S}$ becomes zero at an angle $\theta _* $
$~(0 < \theta _* < \pi/2 )$.
Figure \ref{fig:0} also shows the radial component of magnetic field
$B_{\hat r} \propto G_{S},_\theta $ 
and radial current $4 \pi \alpha j_{\hat r} $ $ \propto  G_{S} ^2 B_{\hat r}$,
where eqs. (\ref{Bcomp}) and (\ref{eqn:current}) are used. 
We denote the angles $\theta_{n}~(n=1,2)$ as the root of $B_{\hat r}=0$.
The direction of $ j_{\hat r} ~(\propto B_{\hat r})$
changes at $\theta_{1}$ in a dipole-dominated field ($0 \le  \xi < 1/3 $), 
whereas it changes twice at $\theta_{1}$ and $\theta_{2}$
in a quadrupole-dominated field ($\xi \ge  1/3 $).
  In this paper, we consider two kinds of models
by restricting the current-flowing region
when the model (\ref{powerlaw}) is applied. 
In the 'whole flowing-model', eq.(\ref{powerlaw}) is 
applied everywhere, irrespective of the sign of $G$.
The current flows in the magnetosphere along a constant line of $G$, 
inwards or outwards at the surface as shown in Fig.\ref{fig:0}. 
The toroidal field $B_{\hat \phi}$ is positive definite
for the parameter $0 \le  \xi < 1/3 $, whereas it changes 
the direction of a border line
started from $\theta_{*}$ on the surface for $\xi \ge  1/3 $.
This current model provides a globally twisted magnetosphere.
Another model is relevant to the quadrupole-like configuration only
($ \xi \ge  1/3 $).
The current flowing region in the magnetosphere is limited by
applying eq.(\ref{powerlaw}) only to a smaller region with 
magnetic field lines starting from  
$0 < \theta < \theta _* ~(< \pi/2)$ on the surface,
that is, a negative region of $G$.
In this 'partially flowing-model', there is a small circuit
started from $\theta_2 < \theta < \theta _* $,
and back to the polar region 
$0 < \theta < \theta _2 $ on the surface, as shown in Fig. \ref{fig:0}.
This model describes a locally twisted magnetosphere. 

  \subsection{Helicity and energy}
\begin{figure}
\centering
\includegraphics[scale=1.20]{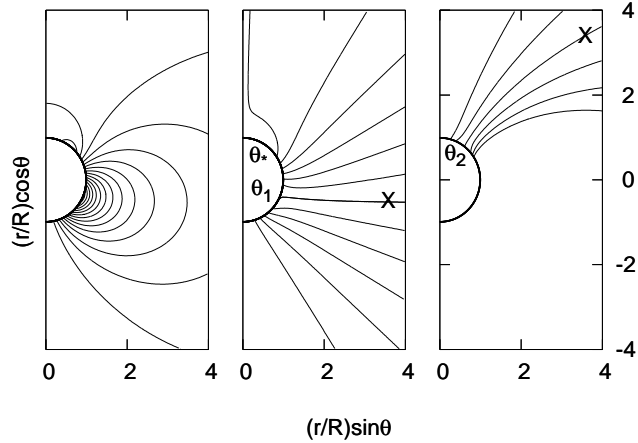}
\caption{
Magnetic field lines for a dipole--quadrupole mixture
in the case of $ \xi >1/3 $.
The surface condition is given in the right panel of Fig.\ref{fig:0}. 
The left panel shows the potential field lines.
The middle and right panels show 
open field lines with the same surface condition.
In the partially open configuration, 
the field lines with $G>0$ are relevant to the opening.
There are still closed lines originating from
$0 < \theta <\theta_*$ on the surface, but they
are not displayed in the middle panel for simplicity. 
The right panel shows partially open field lines with $G<0$,
and closed lines originating from
$\theta_* <\theta <\pi$ are not displayed.
The current sheet is located on a line with a $\times$ symbol,
although this is not necessary for this work.
}
\label{fig:2}
\end{figure}

Two integrals, those of magnetic helicity and energy, are useful to 
characterize the equilibrium solution of the magnetospheres.
Magnetic helicity represents a global property of magnetic fields,
and is obtained by integrating the product of
two vectors, namely, ${\vec A}$ and
 $ {\vec B} (= {\vec \nabla} \times {\vec A})$.
A quantity $H_{\rm R} $, which can be 
defined by the difference between the magnetic helicity of the force-free field
and that of the potential field with the same surface boundary condition.
The total relative helicity in the exterior $(r \ge R)$ is given by
\begin{equation}
H_{\rm R} 
= 4\pi\int_{ r\ge R}  \frac{ GS}{\alpha^{2}}
\frac{dr d\theta }{\sin \theta} .
\label{eqnHR}
\end{equation}
Magnetic energy stored in the force-free magnetosphere is also 
given by integrating over the 3-dimensional volume: 
\begin{equation}
E_{\rm EM} 
= \frac{1}{4}\int_{ r\ge R} \left[ \left( \alpha  
\frac{\partial G}{\partial r} \right)^2
+ \left( \frac{1}{r}\frac{\partial G}{\partial \theta} \right)^2
+\left(\frac{S}{\alpha} 
\right)^2  \right] 
\frac{dr d\theta }{\sin \theta} .
\label{eqn:Eng}
\end{equation}
 The numerical results of $H_{\rm R} $ and $E_{\rm EM}$,
which depends on the twist, is given in the next section.
Here we discuss the energy for two reference configurations.
One is given by that for the potential field for 
a given boundary condition (\ref{SurfG.eqn}). 
It corresponds to the lowest energy and is calculated in terms 
of solutions $g_{l}$$~(l=1,2)$ (eqs.~ (\ref{eqnpuredip})-(\ref{eqnpurequad}) ). 
The energy of a dipolar potential field is $ E_{{\rm d}}= B_{0}^2R^3/3$, and 
that of a quadrupolar potential field is
$ E_{{\rm q}}= 6B_{0}^2R^3/5$ in flat spacetime.
Their values for a relativistic model with $M/R=0.25$ are
numerically calculated as
$ E_{{\rm d}} = 0.74 B_{0}^2R^3$ and $ E_{{\rm q}} = 5.14 B_{0}^2R^3$.
When the boundary field is a dipole--quadrupole 
mixture with a ratio $a_2$, the
energy is given by $E_{0}= E_{{\rm d}}+ a_{2} ^2  E_{{\rm q}}$.
The potential energy of a quadrupole is larger than that of a dipole,
when $a_{2} \ge  (E_{{\rm d}}/ E_{{\rm q}})^{1/2} \approx$ 0.4-0.5,
for the relativistic factor $M/R$. 
Another important energy is $  E_{\rm open}$,
which is energy stored in the open field 
with the same boundary condition (\ref{SurfG.eqn}). 
Suppose the initially closed magnetic field lines of 
a force-free magnetosphere are stretched out to infinity by some artificial means,
keeping the same surface condition;
energy is increased in the new state.
When the energy $E_{\rm EM} $ of a force-free magnetosphere 
exceeds $  E_{\rm open}$,
an open field configuration is preferable in energy.
Such a state with $E_{\rm EM} >  E_{\rm open}$
may be related to the abrupt transition with mass ejection.
Figure \ref{fig:2} schematically demonstrates
the open field configuration.
The quadrupole component is slightly larger than the dipolar one
at the stellar surface, i.e., a case of $ \xi >1/3 $.
The surface condition of this example is
described in the right panel of Fig. \ref{fig:0}.
Corresponding to the surface boundary condition, 
there are two families of closed field lines:
They are either lines with $G>0$ or those with $G<0$.
One of closed field lines becomes open
under the partially open configuration, whereas
both closed lines are open in a fully open configuration.
In the latter, larger energy is associated with the transition.
We only consider a partially open configuration in this paper.
The middle panel of Fig. \ref{fig:2} shows 
the closed field lines of $G>0$ are opened, and
the right panel shows those of $G<0$ are opened.

%
 The method to calculate $E_{\rm open}$ is discussed by 
\citep[e.g.,][]{1993ApJ...410..412L}.
Here, we briefly explain it
to give an example in the case of $ \xi > 1/3$.
We modify the boundary condition (\ref{SurfG.eqn}) at the surface as
\begin{eqnarray}
G^{*} _{S}(\theta) &= & G_{S}(\theta) 
~~~~~~~~~~~~~~( 0 \le \theta \le \theta_{1} ),
\nonumber
\\
G^{*} _{S}(\theta) 
&=& 2G^{*} _{S}(\theta_{1}) -G_{S}(\theta)
~~( \theta_{1} < \theta \le \pi ).
\end{eqnarray}
In the top panel of Fig.\ref{fig:0}, $G_{S}(\theta)$ is plotted by a solid 
line, and $ G^{*} _{S}(\theta) $ by a dotted one.
As inferred from the figure, it is necessary to contain the monopolar function
$1-\cos\theta$ to express $G_{S}(\theta)$.
By solving eq.(\ref{eqn:FF}) with $S=0$ and 
surface boundary condition $G_{S}(\theta)$,
we have a quadrupolar potential field, which is
shown in the left panel of Fig.\ref{fig:2}.
By replacing it with $ G^{*} _{S}(\theta) $, a partially
open field solution $G_{\rm open}$ is obtained.
All the field lines with $G>0$ extend to infinity.
The magnetic configuration is shown in the middle panel of Fig.\ref{fig:2}.
The solution with $ G^{*} _{S}(\theta) $ is unphysical since it contains 
a magnetic monopole charge. The desired solution is obtained by reversing 
its direction only on those lines starting from a region 
of $\theta_{1}  < \theta \le \pi$ on the surface. 
The magnetic energy is unchanged by this sign-flipping,
and may be calculated for the solution $G_{\rm open}$.
Here is a remark on the temporary transition of a twisted magnetosphere.
A field line originating from polar angle $\theta_{1}$ on 
the surface in Fig.\ref{fig:2} corresponds to the current sheet, 
which separates regions of opposite magnetic polarity. 
The open field is strict poloidal, with $B_{\hat\phi} = 0$,
although the force-free field is twisted with $B_{\hat\phi}\ne 0$.
A finite twist is assumed to propagate to infinity along
open field lines. 
The same method is applied to opening
closed lines starting from a region $0 < \theta < \theta_{2}$
on the surface in Fig.\ref{fig:2}, although
it is enough to consider the boundary modification 
with respect to $\theta_{1}$ in the case of $\xi <1/3$. 
Partially open field energy is in general different with respect to 
the opening angle $\theta_{n}$.
The energy regarding $\theta_{1}$ is larger than that for $\theta_{2}$.

\section{Numerical results}
\subsection{A sequence of solutions}
\begin{figure}
\centering
\includegraphics[scale=0.2]{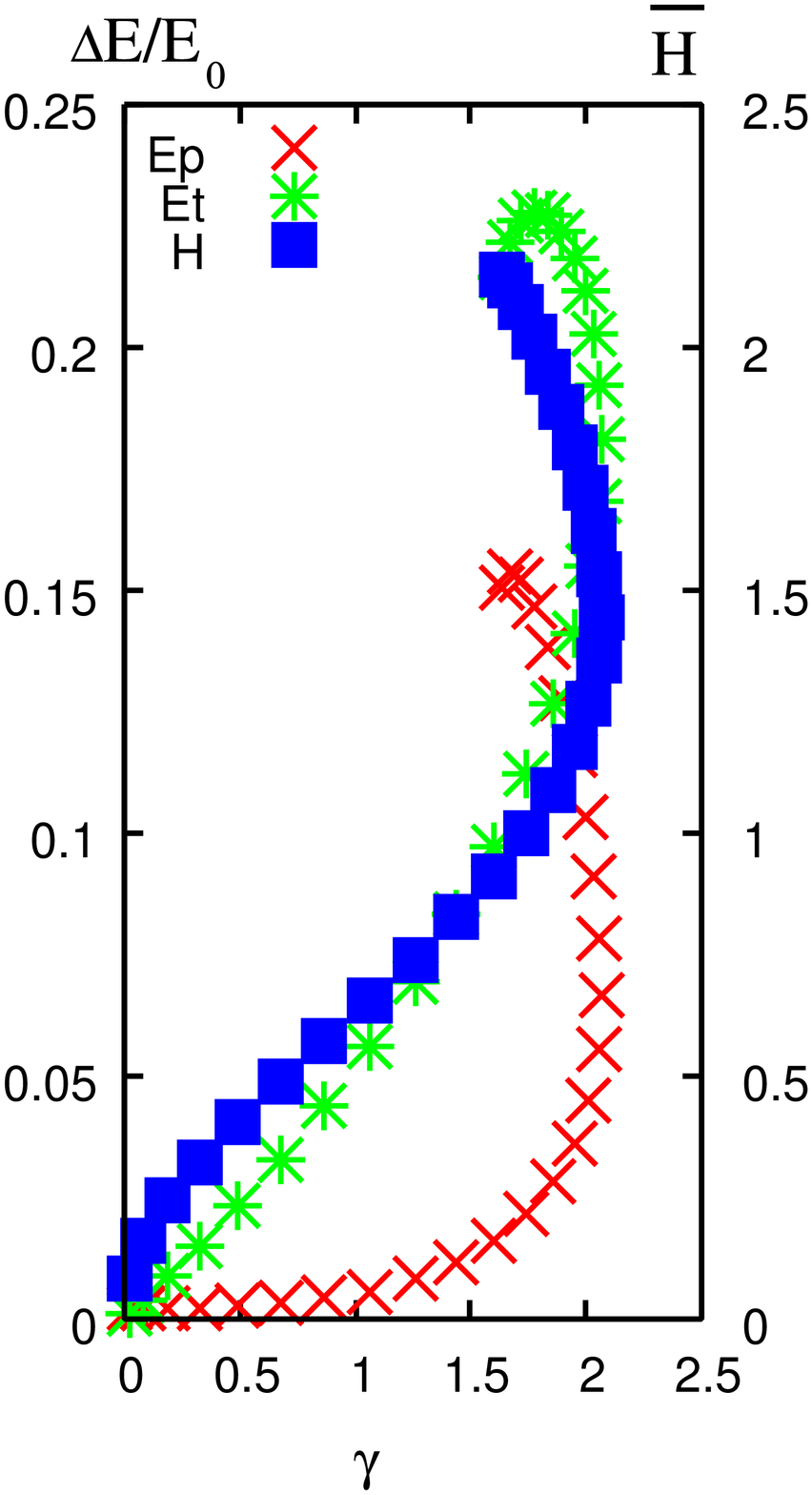}
\includegraphics[scale=0.2]{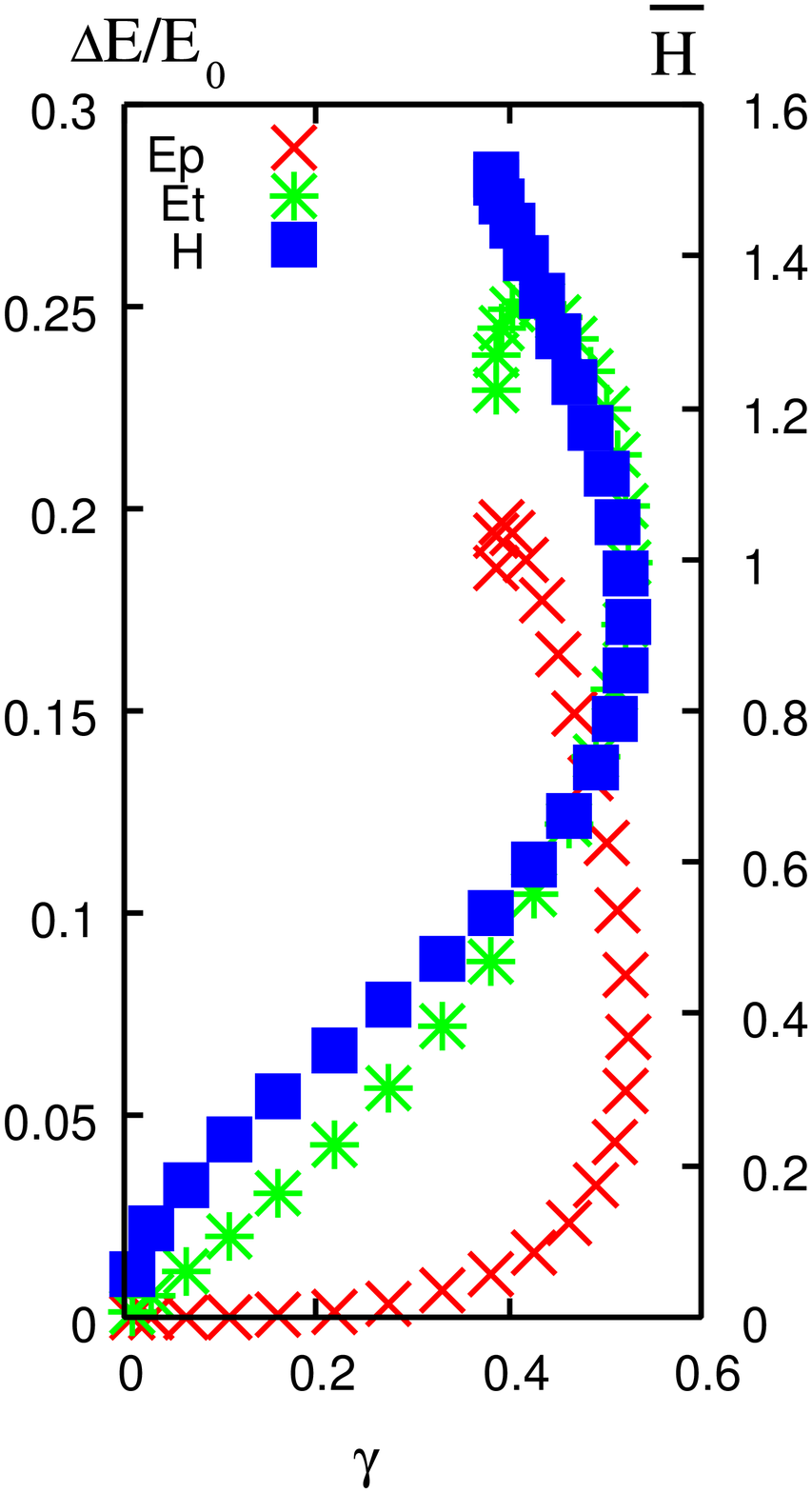}
\includegraphics[scale=0.2]{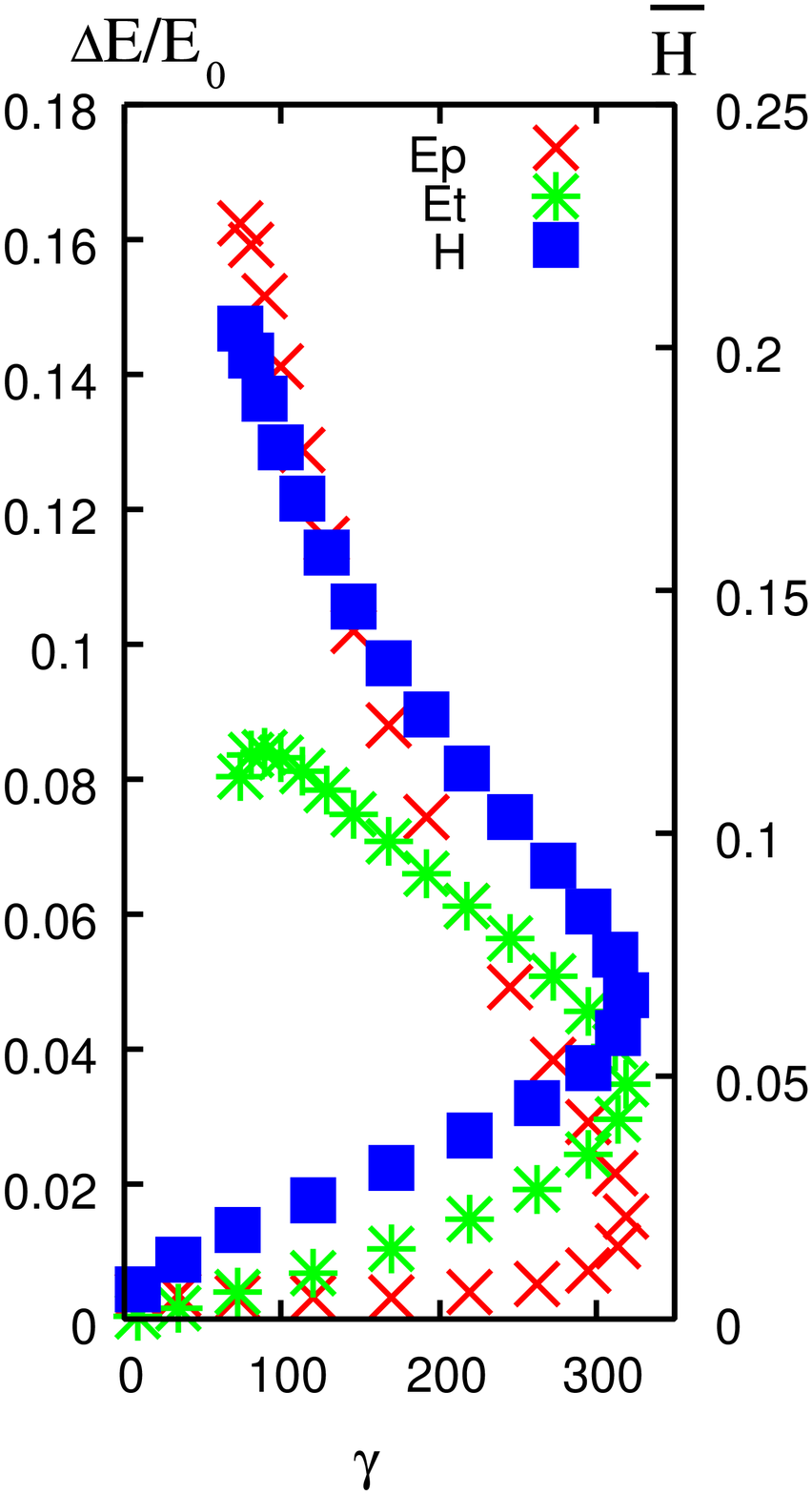} 
\caption{Increase in magnetic energy $\Delta E/E_0$ 
from the potential dipole field is shown on the left axis and
total relative helicity ${\bar H} \equiv H_{\rm R} /(4\pi E_0 R)$
is shown on the right axis.
The poloidal component of the energy is denoted by crosses, 
the toroidal component by asterisks, and helicity by squares.
The horizontal axis denotes 
the dimensionless value $\gamma B_0^4 R^{10} $.
Results for the whole current-flowing model in a flat spacetime
are shown in the left ($a_2=0.2$) and
middle panels ($a_2=1$).
The right panel ($a_2=1$) shows the results for a partially current-flowing model.
}
\label{fig:3}
\end{figure}

\begin{figure}
\centering
\includegraphics[scale=0.2]{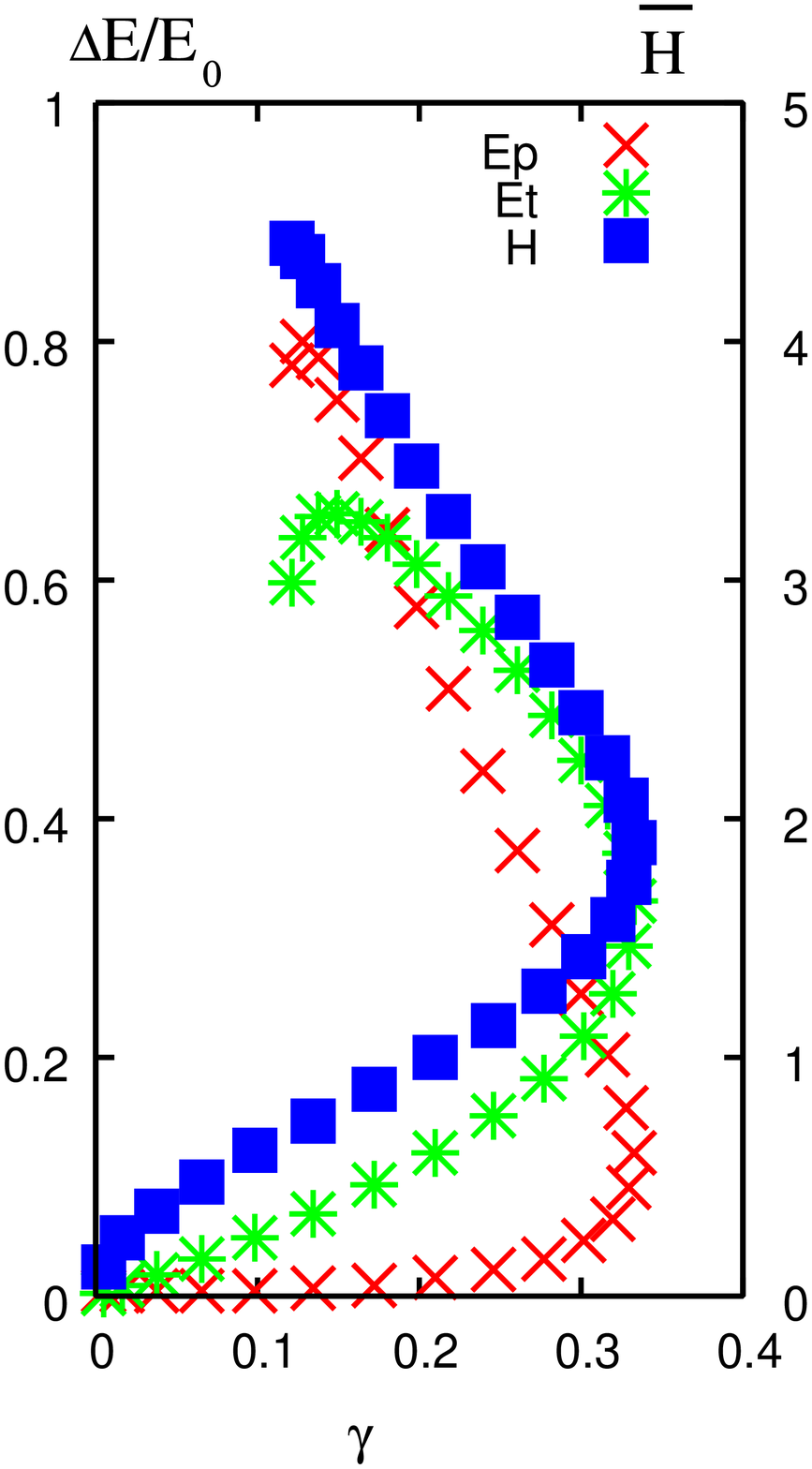}
\includegraphics[scale=0.2]{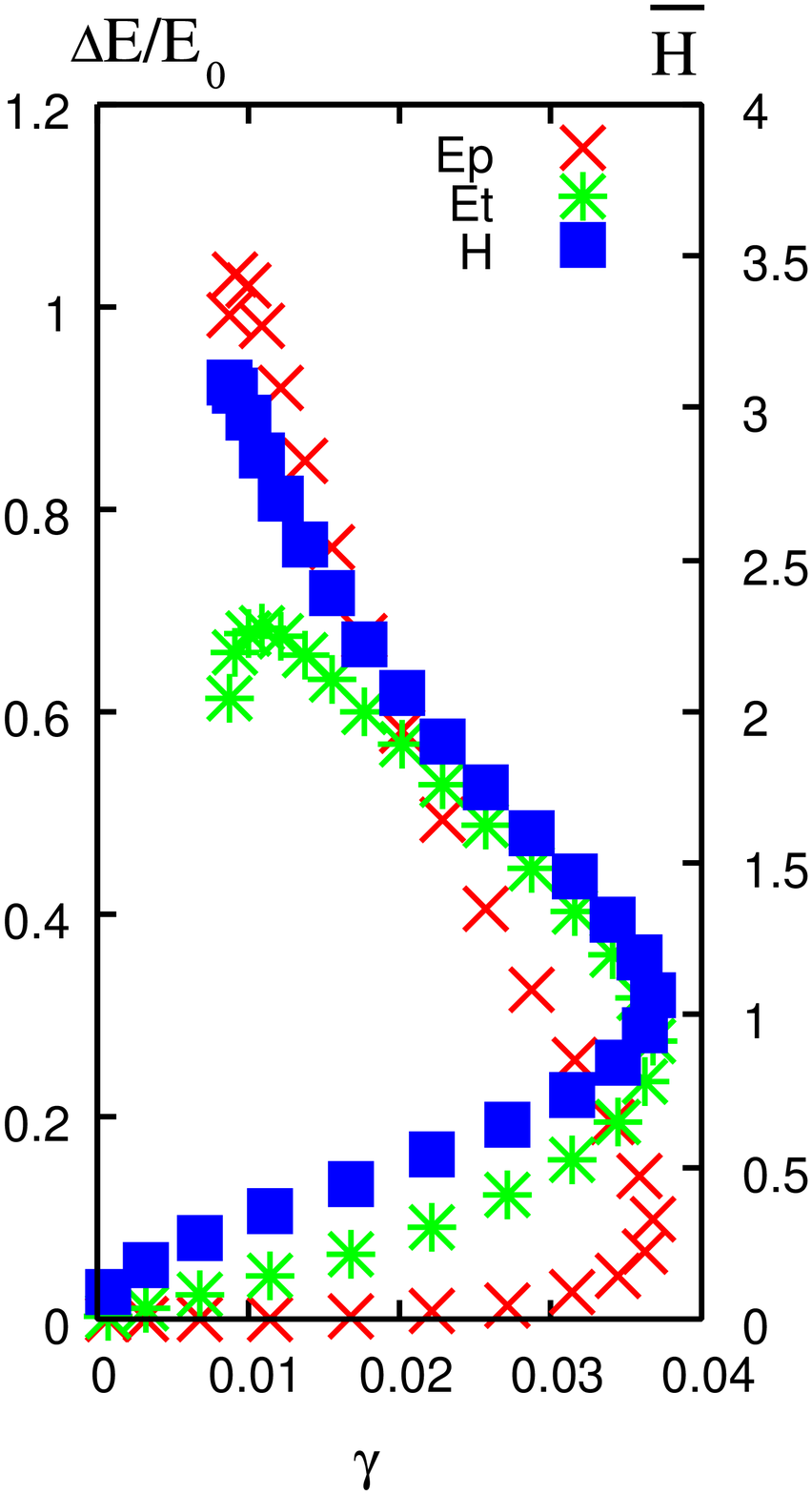}
\includegraphics[scale=0.2]{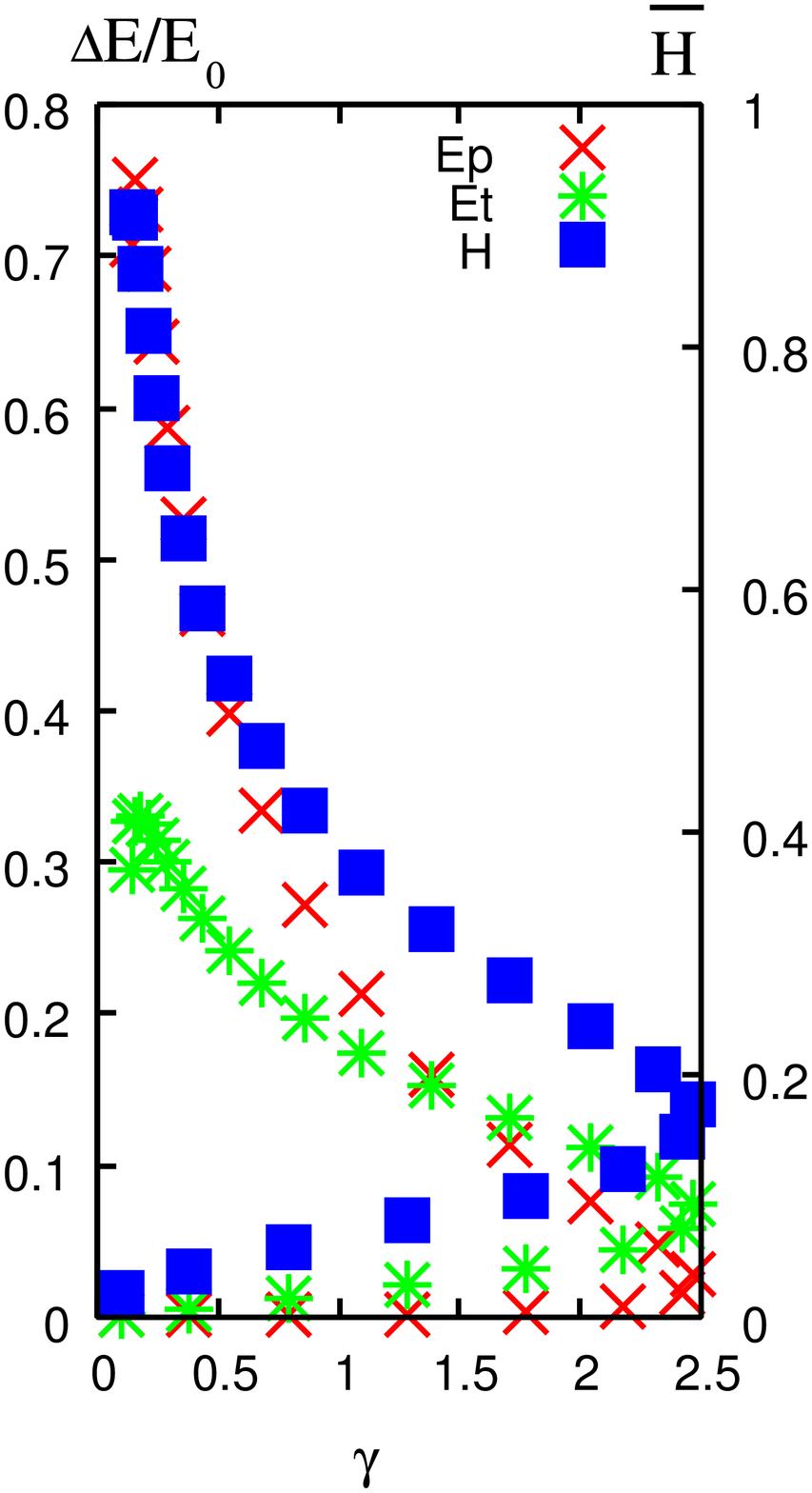} 
\caption{This figure reflects the same conditions as in Fig. \ref{fig:3}, but for
models with a relativistic factor $M/R=0.25$.
}
\label{fig:4}
\end{figure}

A sequence of magnetospheres is numerically constructed
for a fixed boundary condition with a ratio $a_2$
and a relativistic factor $M/R$.
We start with a potential field solution, and follow the structure change 
by increasing the azimuthal magnetic flux or 
helicity, which is used as the degree of twist, and the constant $\gamma$ 
is posteriorly determined.
Thus, both magnetic energy $E_{\rm EM}$ and relative helicity $H_{\rm R}$
are multi-valued functions of $\gamma$.
The increase in magnetic energy and the relative helicity 
for some examples are shown in Figs. \ref{fig:3} and \ref{fig:4}. 
The results in flat spacetime ($M/R =0$) are given in Fig. \ref{fig:3}, and 
relativistic models ($M/R =0.25$) are given in Fig. \ref{fig:4}.
Left and middle panels are results for a whole current-flowing model,
while the right panel corresponds to those for a partially current-flowing model.
For a better understanding of the mechanism, the energy 
difference $\Delta E(=E_{\rm EM}-E_{0})$ is divided into 
$\Delta E =\Delta E_{t}+\Delta E_{p}$, comprising
that of the toroidal component and that of the poloidal component.

The general tendency is the same in all models.
There are two branches in the curves of $\Delta E_{p}$,  $\Delta E_{t}$
and $H_{\rm R}$, as functions of $\gamma$. 
In the lower branch, which corresponds to a weak toroidal magnetic field, 
the energy $\Delta E_{t}$ increases monotonically
with the strength parameter $\gamma$. 
The helicity $H_{\rm R}$ also increases, but
the poloidal energy is almost constant, $\Delta E_{p} \approx 0$. 
There is a maximum of $\gamma$, and after passing the turning point,
$\Delta E_{p}$ drastically increases in the upper branch.
This means a significant structural change from 
that in the potential field. This conclusion is confirmed later.
The curve of $\Delta E_{p}$ or $\Delta E_{t}$
in Figs. \ref{fig:3} and \ref{fig:4} no longer increases,
but curls into a limiting point by further twisting.
This behavior is similar to that often appearing
near a critical point in nonlinear dynamics.
\citet{2004ApJ...606.1210F,2006ApJ...644..575Z,2012ApJ...755...78Z} 
demonstrated the detailed behavior with the current model
in a flat spacetime. A careful treatment is necessary by changing
the parameter near the endpoint.
In this research, we do not resolve the endpoint of the sequence, 
but the maximum value in energy or helicity
is unchanged, even further exploring the termination.
The general property along the sequence is the same whether or not
the relativistic effect is taken into account, but
the maximum values of $\Delta E_{p}$, $\Delta E_{t}$, and $H_{\rm R}$
in Figs. \ref{fig:3} and \ref{fig:4} are quite different.
We normalize the energy excess $ \Delta E$ 
by potential field energy $E_{0}$, 
which depends on magnetic field strength $B_{0}$, 
a ratio $a_{2}$, stellar radius $R$ and relativistic factor $M/R$.
The ratio $\Delta E_{p}/E_{0}$ or $\Delta E_{t}/E_{0}$ 
depends on $a_{2}$ and $M/R$, and clarify the effects of
magnetic field configuration and general relativity.  
The maximum value by the relativistic treatment increases
by a factor of 2 to 5 for fixed $a_{2}$. 
This increase caused by curved space is already demonstrated
for the model with $a_{2}=0$ \citep{2017MNRAS.468.2011K}.


\subsection{Magnetic field configuration}

\begin{figure}
\centering
\includegraphics[scale=1.0]{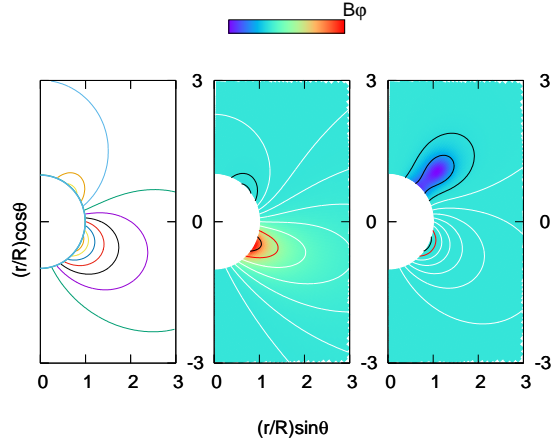}
\caption{Contour lines of magnetic function $G$ and color contour 
of $B_{\hat{\phi}}$ in the $r$-$\theta$ plane for a model with
$a_2 =1$ and $M/R=0$.
The left panel shows the potential field, and
the middle panel shows the highly twisted state with
a whole current-flowing model.
The right panel shows results for a partially flowing model.
}
\label{fig:5}
\end{figure}

\begin{figure}
\centering
\includegraphics[scale=1.0]{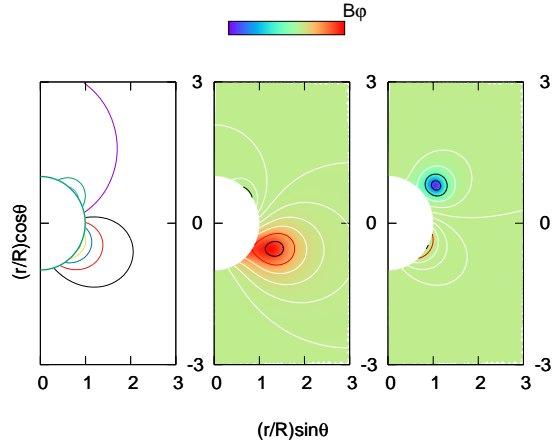}
\caption{The same as Fig. \ref{fig:5}, but for the model with a
relativistic factor $M/R =0.25$.
}
\label{fig:6}
\end{figure}

We present the results of a flat spacetime treatment ($M/R=0$).
Figure \ref{fig:5} shows the magnetic function $G$ with contour lines
and toroidal magnetic field $B_{\hat \phi}$ with colors in the $r$-$\theta$ 
plane. Only the interior part is shown, since the field in the
exterior approaches a vacuum solution since $S \propto G^3 \to 0$, and
the outer part is unchanged.
The surface field is given by a quadrupole dominated field with $a_{2}=1$.
The left panel shows potential field ($B_{\hat \phi}=0$), and
the others show the structure for a highly twisted model.
The middle panel shows the result of the whole flowing-model,
and the right shows that for a partially flowing-model.
They correspond to the end points along a sequence of middle 
and right panels in Fig. \ref{fig:3}.
In comparison with the potential field, 
magnetic lines are stretched outwards by a flowing current.
The maximum of the toroidal field $B_{\hat \phi}$ is located at latitude 
$\theta \approx \theta_1 $ in the middle panel.
This is related to the maximum of the function 
$|G_{S}(\theta)|$ (eq.(\ref{SurfG.eqn})) at surface,
since $B_{\hat \phi} \propto G^3$ in our current model.
The magnetic structure in the northern hemisphere is almost unchanged
in a globally twisted model, since most current flow 
is confined to a region $\pi/2 < \theta \approx \theta_1 < \pi $.
However, an elongated field structure appears in the northern hemisphere
in a constrained current-flow model as shown in the right panel.
Numerical results show that it is possible to sustain strong $B_{\hat \phi}$, 
and that a flux rope is formed.
The center is located at $ (r/R, \theta ) \approx (1.5, \theta_{2} )$.
Note that the flux rope is not formed even in the highly twisted model
with a whole current-flowing model as shown in the middle panel.
Results for the relativistic treatment are given in Fig. \ref{fig:6}.
Formation of a flux rope is more evident in relativistic models.
There is a sharp peak of $B_{\hat \phi}$ in  
the highly twisted state, irrespective of the current flow model.
The configuration shown in the middle and right panels in Fig. \ref{fig:6}
corresponds to the end point along a sequence of middle 
and right panels in Fig. \ref{fig:4}.
Among these highly twisted states shown in Figs. \ref{fig:5} and \ref{fig:6},
there is no flux-rope structure only in the middle panel of Fig. \ref{fig:5}.
The model is also different in the energy increase $\Delta E $
considered in Figs. \ref{fig:3} and \ref{fig:4}:
$\Delta E_{t} >\Delta E_{p}$ in this model.
In other models at the end point, we have $\Delta E_{p} >\Delta E_{t}$.
The flux-rope contains a strong toroidal magnetic field, but
a strong poloidal component is also needed to support the toroidal component.
Thus, the flux-rope formation is associated with
a significant structural change from a potential magnetic field in vacuum.
This property is also numerically supported by 
previous research \citep{2017MNRAS.468.2011K,2018MNRAS.475.5290K}.
As inferred from Figs. \ref{fig:3} and \ref{fig:4},
the flux-rope is associated in the upper branch.
So the state is able to store more energy, and
may be meta-stable.
Some elaborated methods are necessary to calculate such a state.
There are some numerical results in which 
the flux rope is extended to several times the stellar radius, 
$r_{c}/R \sim  5$
in the models by \citet{2015MNRAS.447.2821P,2018MNRAS.474..625A}.
As an input parameter to construct a force-free magnetosphere, 
they fix the value of the critical radius $r_c$
which is defined as the radial extent of the current flow region on the
equatorial plane.
Our results show that the location is not far from the central star.
The center of the loops is within $r_{c}/R < 1.5$.

\subsection{General relativistic confinement}

\begin{figure}
\centering
\includegraphics[scale=1.0]{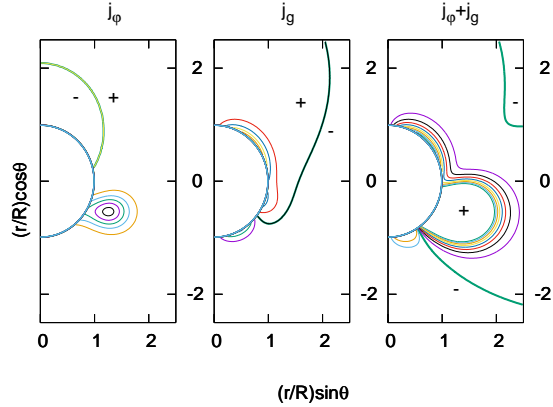}
\caption{
Contour azimuthal currents in eq.(\ref{eqn:FFNR})
for the model corresponding to the middle panel of Fig. \ref{fig:6}.
The thick curves denote the zero level, and
the $+$ or $-$ symbols denote the sign of the
current at these points.
The left panel shows that $j_{\hat \phi}$ has
a sharp peak at $(r/R, \theta) \approx (1.2, 120^{\circ} )$, which
corresponds to the center of the flux rope in Fig. \ref{fig:6}. 
The middle panel shows a 'general-relativity-induced current' $j_{g}$,
which is in the opposite direction of $j_{\hat \phi}$ around each polar region.
The right panel shows a sum of $j_{\hat \phi}$ and $j_{g}$.
There is still a sharp peak around the center of each flux rope, 
but their higher level contours are omitted. 
}
\label{fig:7}
\end{figure}

\begin{figure}
\centering
\includegraphics[scale=1.0]{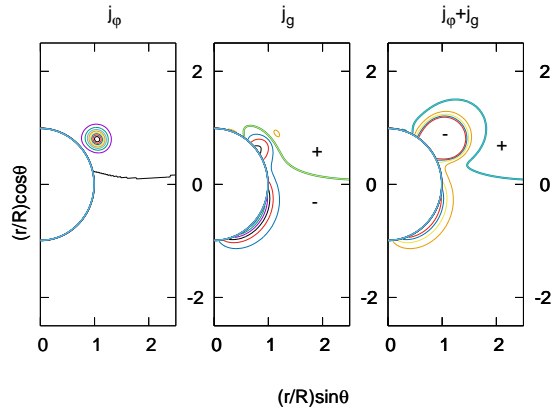}
\caption{
This figure reflects the same conditions as Fig.\ref{fig:7}, but for the model corresponding to 
the right panel of Fig. \ref{fig:6}.
A sharp peak of $j_{\hat \phi}$ at $(r/R, \theta) \approx (1.2, 60^{\circ} )$
in the left panel corresponds to the center of the flux rope in Fig. \ref{fig:6}. 
In the right panel, there is still a sharp peak of $j_{\hat \phi}+j_{g}$ 
around the center of the flux rope, 
but their higher level contours are omitted. 
}
\label{fig:8}
\end{figure}

In this subsection, we discuss the general relativistic effect
on the flux rope formation.
The effect is so far studied through numerical calculations.
Here, we explain it by an interpretation of
additional current $j_{g}$ in a flat spacetime formulation,
eq.(\ref{eqn:FFNR}).
The 'general-relativity-induced current' $ j_{g}$ is not
so large, and the ratio to true current $j_{\hat \phi}$ 
is $j_g / j_{\hat \phi} \sim 0.1$ in its magnitude.
However, this small term is crucial in the exterior of a flux rope.
Figure \ref{fig:7} demonstrates the current distribution for
$j_{\hat \phi}$ and $ j_{g}$.
The model corresponds to a highly twisted state
for the whole current flowing model with $M/R=0.25$
(middle panel of Fig. \ref{fig:6}).
There is a sharp maximum of $j_{\hat \phi} ~(>0)$.
The position corresponds to a center of the flux rope.
In the distribution of $j_{\hat \phi}+j_{g}$ (right panel),
the sharp peak-structure is still unchanged.
However, the outer part is covered by an effectively negative current.
This opposite current-flow supports the flux-rope formation.
In Fig. \ref{fig:8}, we also show another example
for a partially current-flowing model with $M/R=0.25$
(right panel of Fig. \ref{fig:6}).
In this case, there is a strong current 
$j_{\hat \phi} \le 0$ in the northern hemisphere.
The rope is surrounded with a positive current $ j_{g}$ 
produced by 'gravity.'
In both models, the term $ j_{g}$ 
is in an opposite direction to the true current $j_{\hat \phi}$.
The mechanism may be understood as follows.
Supposed that a flux rope with $ j_{\hat \phi} >0$ is formed.
The magnetic function $G$ is a maximum in our model, and
it outwardly decreases, i.e., $G_{, r} <0$.
Equation (\ref{eqn:GRj}) provides $ j_{g} <0$ in the outer part
of the rope. On the other hand, in the case of $ j_{\hat \phi} <0$,
the center of the flux rope is the minimum
of the magnetic function, and therefore $G_{, r} >0$.
Equation (\ref{eqn:GRj}) provides $ j_{g} >0$ in the outer part.
Whichever the direction of a current flow in the rope is, 
a counter-flow of effective current is produced in the exterior,
and it holds the flux rope.

\subsection{Energy}

\begin{figure}
\centering
\includegraphics[scale=1.0]{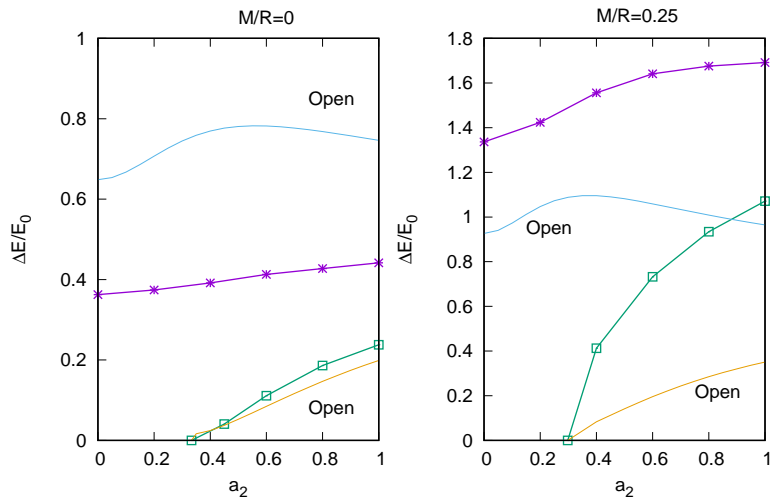}
\caption{
Maximum energy increment $\Delta E/E_0$ 
as a function of the quadrupole component $a_2$.
The left panel shows results with $M/R =0$, while
the right shows $M/R =0.25$.
Maximum energy with the whole current-flowing model
is denoted by asterisks, and that with
a partially current-flowing model is denoted by squares.
The latter is possible for the range $a_2 >g_{2}/(3g_{1}) \approx 1/3$.
A curve labeled 'open' denotes the partially open field energy 
$\Delta E_{\rm open}/E_0 =(E_{\rm open} -E_{0})/E_0$.
}
\label{fig:9}
\end{figure}

%
  As shown in Figs. \ref{fig:3} and \ref{fig:4},
the energy deposited in a magnetosphere increases with twisting 
for a given surface condition.
Figure \ref{fig:9} shows the maximum increment 
$\Delta E_{\rm max}/E_{0} \equiv E_{\rm max}/E_{0}-1$,
as a function of a ratio $a_{2}$, where 
$E_0$$(= E_{{\rm d}}+ a_{2} ^2  E_{{\rm q}})$ is the potential field energy.
We also plot the energy difference  
$\Delta E_{\rm open}/E_{0}  \equiv (E_{\rm open}-E_{0})/E_{0}$ 
between the partially open energy $E_{\rm open}$ and potential energy $E_0$.
Results with $M/R=0$ are shown in the left panel. 
Both $\Delta E_{\rm max}/E_{0} $ and $\Delta E_{\rm open}/E_{0}$ 
are less sensitive to $a_{2}$
in the whole current flowing magnetosphere:
$\Delta E_{\rm max}/E_{0} \sim$ 0.34-0.44 and 
$\Delta E_{\rm open}/E_{0} \sim$ 0.66-0.78 for $0\le a_{2}\le 1$. 
It is known that $\Delta E_{\rm max}/E_{0} =0.34 <$ 
 $\Delta E_{\rm open}/E_{0}  =0.66$ 
with a purely dipole boundary in flat spacetime\citep{2004ApJ...606.1210F}. 
It is impossible that the maximum energy exceeds the open field energy
even by including the quadrupole component in the flat model.
The situation however changes, i.e.,  $E_{\rm max} >E_{\rm open}$  
In the partially flowing model, which is relevant to $a_{2} > 1/3$.
The magnetic field configuration with
a partially open energy $E_{\rm open}$ corresponds to
the right panel in Fig. \ref{fig:2}.
The relevant energy is smaller than that shown in the middle panel, 
since the spatial region is limited.
Figure \ref{fig:9} shows that
ratios $\Delta E_{\rm max}/E_{0}$ and $\Delta E_{\rm open}/E_{0}$ 
monotonically increase with $a_{2} $, since
the current-flowing volume increases. 
The maximum energy is slightly larger than the open field energy,
and the excess is approximately 10\%
( $E_{\rm max}\approx 1.1 E_{\rm open}$) in flat models. 

Results in the framework of general relativity
are given in the right panel of Fig. \ref{fig:9}.
The maximum excess $\Delta E_{\rm max}/E_{0}~$ ($\sim$ 1.34-1.69) 
in a model with $M/R=0.25$ 
is significantly larger than that in a model with $M/R=0$. 
The large increase is closely related to the flux rope formation
in a relativistic model \citep{2017MNRAS.468.2011K,2018MNRAS.475.5290K}.
Furthermore, the maximum energy exceeds 
$E_{\rm open}$ in any model with $M/R=0.25$. 
The excess is approximately 20--40\%
($E_{\rm max}\approx$ 1.2-1.4 $E_{\rm open}$). 
In the partially current-flowing model, 
fractional excess further increases by up to approximately 30--50\%
($E_{\rm max}\approx$ 1.3-1.5 $E_{\rm open}$). 
The energy in a magnetic field configuration containing an evident flux rope,
as shown in Figs. \ref{fig:5} and \ref{fig:6}, exceeds the open field energy.
Substantial amount of energy is stored in the rope, and
is likely to eject in dynamical transition.

\section{Discussion}
In this paper, we have studied the energy storage in a relativistic 
force-free magnetosphere with a quadrupole component on the surface.
The component is introduced to examine the effect of local irregularity 
of the magnetic fields, since a purely dipolar field is the ideal case, 
and higher multi-poles are likely to be involved on the magnetar surface.
We calculated a sequence of equilibrium models 
whose evolution may be characterized by increasing the helicity stored 
in the magnetosphere.
The magnetic energy also increases along the sequence, and
exceeds the open-field one in some cases.
A state with $E_{\rm EM} > E_{\rm open}$ is regarded as a metastable state 
in its energetics. It may be changed to a lower energy state
through an intermediate, open magnetic-field structure.
During the transition, the magnetic flux rope, 
in which plasma is also contained, is ejected.
This dynamical event may be related to an observed magnetar flare.
However, the amount of energy in the catastrophic transition is
ambiguous, since it depends on the stability of higher energy states.
That is, the excess $\Delta E =E_{\rm EM} - E_{\rm open}$ is
almost zero when some kinds of instabilities set in soon 
after reaching a state with $E_{\rm open}$.
On the other hand, $\Delta E $ is further built up by remaining 
in a metastable state longer, and a huge amount of energy is possibly released.
Such a transition requires a dynamical calculation for its solution
\citep[e.g.,][as resistive simulation in flat spacetime]
{2012ApJ...746...60L,2013ApJ...774...92P,2014PTEP.2014b3E01K},
which is beyond the scope of the quasi-equilibrium approach used here. 
We discuss the astrophysical observation.
The energy-scale in most magnetar outbursts is 
not as large as the maximum stored in a dipolar magnetosphere,
$E_{\rm EM} \approx 0.1(B_{p})^2R^3$ 
$=10^{46}(B_p/10^{14.5}{\rm G})^2(R/12{\rm km})^3$ erg.
Here the energy is estimated in flat spacetime approximation 
using magnetic field strength $B_p$ at the surface pole and 
stellar radius $R$.
The general relativistic effect increases it by a factor 2-3.
The outburst energy is $10^{41}-10^{42}$ erg\citep{2018MNRAS.474..961C},
such that the fraction is $\approx 10^{-4}$
(see footnote
\footnote{ In giant flares, 
the energies are scaled up to $10^{44}$-$10^{46}$ ergs.
The ratio is $\approx 10^{-2}$, since
their sources (SGR 0526-66, SGR 1900+14, and SGR 1806-20)
have somewhat stronger field strength
$B_p= (5-20)\times 10^{14}$G\citep{2015RPPh...78k6901T}
}). 
Thus, the outburst is a small reconfiguration of the magnetic fields.
One possibility to account for it 
is that the energy is not built up so much beyond a
state with $\sim  E_{\rm open}$. The metastable state is likely to be destroyed. 
The second possibility is a structural change at a smaller scale.
In this paper, we studied it by considering a locally twisted model
with a limited current-flowing region. 
We considered current flow in a localized region produced by a quadrupole field.
We found that the associated energy is naturally reduced, and 
that a state with $ E_{\rm EM} > E_{\rm open} $ 
is possible irrespective of the spacetimes.
In a flat treatment, the possibility
for $ E_{\rm EM} > E_{\rm open} $ is already explored
\citep{2007ApJ...660.1683W,2012ApJ...750...25W}.
They found greater excess energy in the presence of an exterior potential field
covering the non-potential field.
In our models, which are different from theirs at some points,  
the maximum excess energy is 5\% of the 
potential field energy in a flat spacetime.
The value is significantly increased in a general relativistic model,
up to 70\%.
The general relativistic effect confines the current flow in the vicinity of the star.
The mechanism may be interpreted as a
'general-relativity-induced current,' which
flows in a counter direction so as to cover a flux rope.
The effect is also important for a globally twisted magnetosphere,
when the flows are extended across the entire space. 
The force-free approximation is also considered in the context of 
the pulsar magnetosphere, in which stellar rotation is more important; 
$|j| \ll \rho_{\rm GJ} c \sim \Omega B c$, where $\Omega$ is angular velocity.
The model is recently examined in a general relativistic framework
\citep{2014PhRvD..89h4045R,2016MNRAS.455.3779P,
2016ApJ...833..258G,2017ApJ...851..137G}.
Compared with the estimate based on a flat spacetime treatment,  
total Poynting power increases by 20--60\%. 
The general-relativistic effects are also studied in a vacuum electromagnetic 
field, exterior field of stationarily rotating 
or oscillating relativistic object in the literature
\citep[e.g.,][and the references therein]{1986Ap&SS.120...27M,
2001MNRAS.322..723R,2004MNRAS.348.1388K,
2010MNRAS.408..490M,2017MNRAS.472.3304P};
these researchers found that the general-relativistic correction is of the order 
$G_{\rm N}M/(Rc^2) \sim 0.2-0.3$.
The effects on a static twisted magnetosphere considered 
here and in previous publications 
\citep{2017MNRAS.468.2011K,2018MNRAS.475.5290K}
are found to be substantial, since
a topological change of the magnetic field structures is associated.
Outward eruption of magnetic flux is suppressed,
and relativistic models are 
capable of storing significantly more energy than the 
corresponding potential field energy.
Therefore, general relativistic effects will be taken into account
in magnetar models. The stability and more detailed modeling remain for
future research. 

\section*{Acknowledgements}
This work was supported by JSPS KAKENHI Grant Numbers 
JP26400276 and JP17H06361.

 \bibliographystyle{mnras}
 \bibliography{kojima18mar} 

\end{document}